\documentclass[lettersize,journal]{IEEEtran}
\usepackage{amsmath,amsfonts}
\usepackage{algorithmic}
\usepackage{algorithm}
\usepackage{array}
\usepackage[caption=false,font=normalsize,labelfont=sf,textfont=sf]{subfig}
\usepackage{textcomp}
\usepackage{stfloats}
\usepackage{url}
\usepackage{verbatim}
\usepackage{graphicx}
\usepackage{cite}
\usepackage{xcolor}
\usepackage{tabularx}

\begin{document}

\title{Bridging Neural Networks and Wireless Systems with MIMO-OFDM Semantic Communications}

\author{Hanju~Yoo,~\IEEEmembership{Student Member,~IEEE,}
    Dongha~Choi,~\IEEEmembership{Student Member,~IEEE,}
    Yonghwi~Kim,~\IEEEmembership{Member,~IEEE,}
    Yoontae~Kim,~\IEEEmembership{Student Member,~IEEE,}
    Songkuk~Kim,~\IEEEmembership{Member,~IEEE,}  
    Chan-Byoung~Chae,~\IEEEmembership{Fellow,~IEEE,} and Robert~W.~Heath,~Jr.,~\IEEEmembership{Fellow,~IEEE}%
    \thanks{H. Yoo, D. Choi, Y. Kim, Y. Kim, S. Kim*, and C.-B. Chae* are with the School of Integrated Technology, Yonsei University, Seoul 03722, South Korea (e-mail: \{hanju.yoo, ellijah1030, eric\_kim, kyt0410, songkuk, cbchae\}@yonsei.ac.kr). R. W. Heath, Jr. is with the Department of Electrical and Computer Engineering, University of California San Diego, La Jolla, CA, USA (e-mail: rwheathjr@ucsd.edu). *Co-corresponding authors.}%
    \thanks{This work was in part supported by the Institute of Information \& Communications Technology Planning \& Evaluation (IITP) grant funded by the Korean government (MSIT) (Nos. RS-2024-00428780 and RS-2024-00453301). The work of R. W. Heath, Jr. was supported in part by the National Science Foundation under Grant Nos. NSF-ECCS-243526, NSF-CCF-2435254, and NSF-CNS-2433782, and in part by funds from federal agencies and industry partners as specified in the Resilient \& Intelligent NextG Systems (RINGS) program.}%
    \thanks{Manuscript received January XX, 2025; revised April XX, 2025.}%
}

\markboth{to appear in IEEE Wireless Communications, ~2025}%
{Yoo \MakeLowercase{\textit{et al.}}: Bridging Neural Networks and Wireless Systems with MIMO-OFDM Semantic Communications}

\IEEEpubid{0000--0000/00\$00.00~\copyright~2025 IEEE}

\maketitle

\begin{abstract}

Semantic communications aim to enhance transmission efficiency by jointly optimizing source coding, channel coding, and modulation. While prior research has demonstrated promising performance in simulations, real-world implementations often face significant challenges, including noise variability and nonlinear distortions, leading to performance gaps. This article investigates these challenges in a multiple-input multiple-output (MIMO) and orthogonal frequency-division multiplexing (OFDM)-based semantic communication system, focusing on the practical impacts of power amplifier (PA) nonlinearity and peak-to-average power ratio (PAPR) variations. Our analysis identifies frequency selectivity of the actual channel as a critical factor in performance degradation and demonstrates that targeted mitigation strategies can enable semantic systems to approach theoretical performance. By addressing key limitations in existing designs, we provide actionable insights for advancing semantic communications in practical wireless environments. This work establishes a foundation for bridging the gap between theoretical models and real-world deployment, highlighting essential considerations for system design and optimization.

\end{abstract}



\section{Introduction}
As next-generation wireless networks push toward efficient and intelligent data delivery, conventional digital communication---based on separate source coding, channel coding, and modulation---begin to show their limitations. While classic information-theoretic results~\cite{shannon1949mathematical} support this modular structure, real-world conditions often deviate from the ideal assumptions of large block lengths, stationary channels, and additive white Gaussian noise (AWGN) channel. As a result, practical systems struggle to fully realize the theoretical gains promised by such layered approaches~\cite{polyanskiy2010channel}.

Semantic communications have emerged as a promising alternative~\cite{BeyondBits}. {Semantic communication systems aim to convey the \emph{meaning} of source information rather than merely minimizing symbol-error rates.
Since this objective aligns closely with source encoder optimization, such systems are typically realized through joint source and channel coding using various distortion measures~\cite{bourtsoulatze2019deep,qin2021semantic,tung2021deepwive}.
Semantic communications can also serve as a means to implement goal-oriented systems when the loss functions are carefully designed to align with specific receiver-side tasks~\cite{qin2021semantic}.
The output of a semantic encoder may vary depending on the architecture, producing analog or quantized complex-valued symbols~\cite{bourtsoulatze2019deep}, or even bit streams~\cite{park2024joint, ding2024adaptive}.}



{In this article, we focus on deep learning-based semantic communication systems that directly map input data to complex-valued symbols~\cite{bourtsoulatze2019deep,qin2021semantic,tung2021deepwive}, commonly referred to as deep joint source-channel coding (DeepJSCC).}
This approach has demonstrated improvements over traditional methods in simulations, achieving better performance across metrics ranging from mean squared error to perceptual similarity~\cite{bourtsoulatze2019deep, tung2021deepwive, qin2021semantic}. 
However, their end-to-end design, which bypasses traditional bit-based processing and directly operates on symbols, increases their susceptibility to channel noise and variability.
This raises a question: \textit{Can the theoretical and simulated gains of semantic communications be reliably achieved in real-world wireless scenarios?}

\IEEEpubidadjcol

{
Several prior studies have explored semantic communication prototypes~\cite{yoo2022real, yoo2023role, liu2022real, ding2024adaptive}, but these implementations often exhibit notable performance gaps compared to simulations. The causes of these discrepancies remain understudied, and the practical challenges of deploying semantic communication systems over real wireless channels and hardware platforms are largely unexplored.
This highlights the need for a deeper understanding of the practical constraints affecting semantic communication performance.
}

{
In this article, we investigate the underlying reasons for the performance gap between simulation and prototype results, focusing on the impact of hardware limitations and wireless channel characteristics. To this end, we develop a multiple-input multiple-output (MIMO) and orthogonal frequency division multiplexing (OFDM)-based semantic communication prototype. Unlike typical simulation studies, we account for practical challenges such as power amplifier (PA) nonlinearity and noise variation across subcarriers and spatial streams. Our measurements reveal how these factors contribute to performance degradation, offering a detailed explanation of previously observed gaps and providing insights into the feasibility of semantic communications in real-world systems.
}

{
To the best of our knowledge, this is the first work to validate semantic communication performance using a MIMO-OFDM prototype. Our main contributions are summarized as follows:
}


\begin{itemize} 
    \item We examine challenges encountered in real wireless environments, such as error variations across subcarriers and data streams, to identify the causes of performance deviations between simulations and prototype results.
    \item We demonstrate that semantic communications can achieve near-theoretical performance in real systems when channel variations are properly addressed.
    \item {We validate that semantic communication systems with peak-to-average power ratio (PAPR)-regulation loss~\cite{shao2022semantic, liu2023comprehensive}, which often underperform in simulations without PA nonlinearity, can outperform non-regulated systems in practical scenarios operating within the nonlinear power region.}
    \item We make our source code and hardware setup publicly available\footnote{\url{https://github.com/kmsiapps/semantic-mimo-ofdm}} to provide a foundation for developing semantic communications systems in real-world conditions.
\end{itemize}

\begin{figure*}[t]
    \centering
    \includegraphics[width=\textwidth]{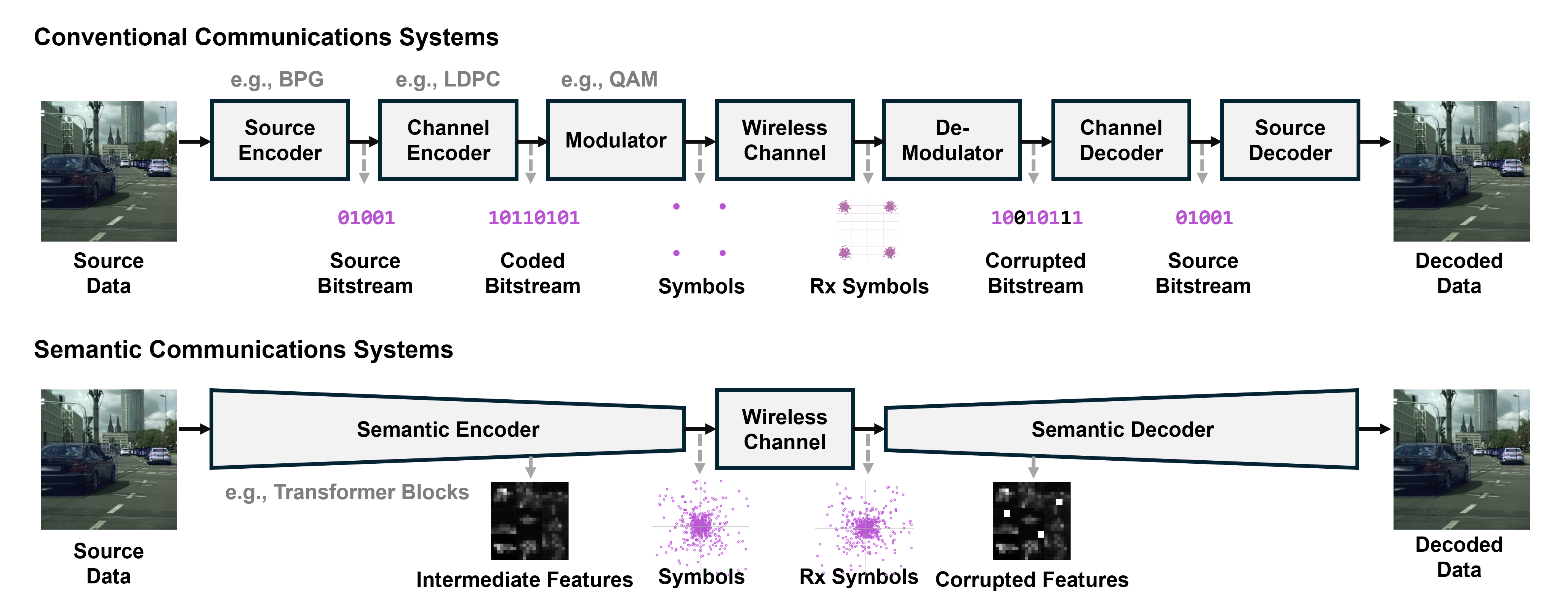}
    \caption{Basic block diagram comparing conventional digital communications and semantic communications system models. Conventional systems rely on separate source coding, channel coding, and modulation stages, whereas semantic communications adopt an end-to-end approach. This approach directly maps source data to wireless symbols, bypassing intermediate bitstreams, and reconstructs the data from noisy symbols. While it enables goal-oriented training, it is more sensitive to channel characteristics.}
    \label{fig:semantic_scheme}
\end{figure*}


\section{An Overview of Semantic Communications}
\label{sec:sc_overview}

\subsection{System Architecture}

In this section, we introduce the typical architecture of semantic communications systems and compare it to conventional communications systems. Fig.~\ref{fig:semantic_scheme} illustrates the basic building blocks of conventional and semantic communications system models. In traditional communication systems, transforming input data into in-phase and quadrature-phase (I/Q) symbols involves multiple stages, including source coding (e.g., JPEG), channel coding (e.g., LDPC), and modulation (e.g., QAM). The output of the source and channel coding blocks is in the form of bits, which are then converted into complex values during the modulation stage.

In contrast, semantic communications systems bypass this bit-based processing by directly mapping input data (e.g., images) to wireless symbols. This method eliminates the need for separate source coding, channel coding, and modulation stages. By jointly optimizing these components with an end-to-end training approach, semantic communications systems enhance efficiency and performance. They consist of an encoder that reduces the dimensionality of the input and a decoder that generates the desired output from the encoded data, typically constructed using deep neural networks.

The encoder in semantic communication systems compresses input data into compact real-valued features, which are paired to form complex symbols for wireless transmission. A differentiable channel layer, typically modeled as AWGN or Rayleigh fading, simulates the physical channel. The impaired symbols are then converted back into corrupted features and processed by the decoder.

The decoder's role is to effectively achieve predefined goals using the corrupted features. These goals can vary, ranging from accurate reconstruction (e.g., image or text reconstruction) to analysis tasks (e.g., object detection or image classification) based on the input data. The extent of achievement is measured using a loss function, which could be a perceptual metric, mean squared error (for reconstruction tasks), or classification accuracy, depending on the system's configuration. As the encoder and decoder jointly learn to achieve these goals despite channel-corrupted symbols, the encoder is trained to generate noise-robust wireless symbols, while the decoder is optimized to efficiently realize the objectives from the encoder-produced symbols.

This end-to-end training process of the semantic communications system, as demonstrated in prior work~\cite{bourtsoulatze2019deep, tung2021deepwive}, has shown improvements in transmission efficiency compared to traditional systems. Transforming features into wireless symbols, however, requires conscientious handling, mainly when those neural network-generated symbols are transmitted over actual wireless channels. We investigate the procedures of symbol transmission and identify the associated challenges in the following sections.


\begin{figure*}[t]
    \centering
    \includegraphics[width=\textwidth]{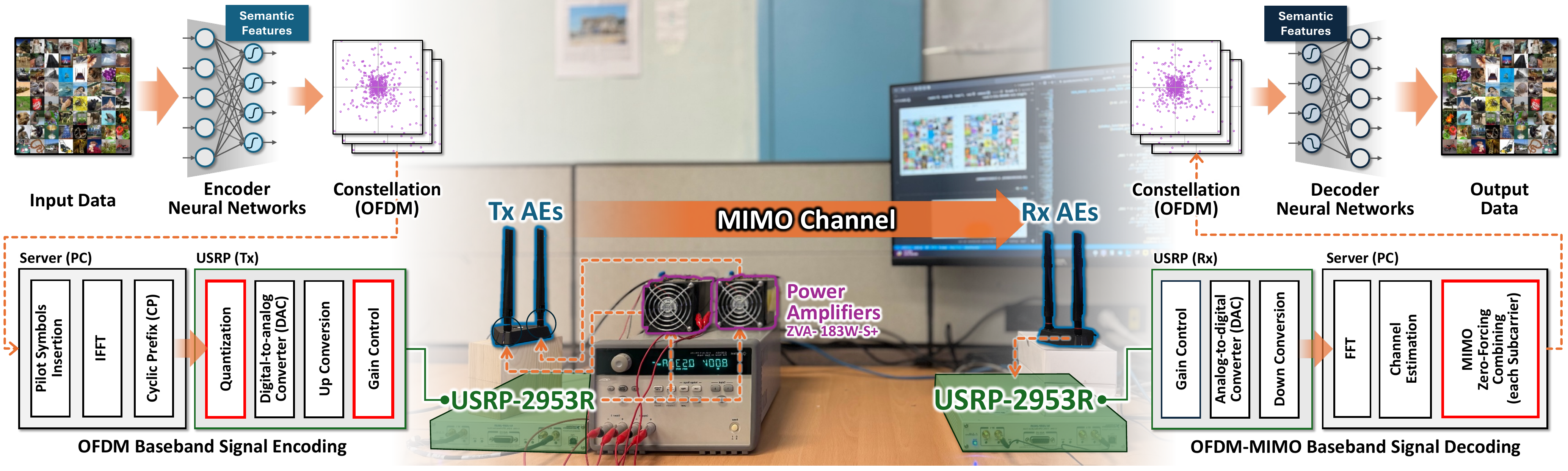}
    \caption{System architecture of a semantic communications system with a MIMO-OFDM prototype setup. Red boxes indicate practical issues arising from the radio hardware and communication channels. Our testbed includes both $2\times2$ and $4\times4$ MIMO configurations.}
    \label{fig:testbed}
\end{figure*}


\subsection{Prior Work}

There have been few implementations of wireless semantic communications prototypes. 
The first effort to implement and validate semantic communications in a wireless environment is documented in~\cite{yoo2022real, yoo2023role}. Using the NI USRP software-defined radio (SDR) device, this work revealed discrepancies between simulated results and real-world measurements in decoded image quality.
 \cite{liu2022real, ding2024adaptive} also validated the performance of semantic communications using the USRP, implementing an OFDM-based wireless transmission system. However, they reported significant performance gaps in the lower transmit power region~\cite{liu2022real} and did not compare favorably to simulated results~\cite{ding2024adaptive}. This article differentiates itself from prior work by analyzing the factors in actual RF hardware and wireless channels that contribute to these performance gaps, and by providing practical insights to guide the design of future semantic systems. Furthermore, all these implementations are based on single-input single-output (SISO) systems, unlike modern communication systems that use multiple transceiver chains.

There have been prior studies addressing practical challenges in implementing semantic communication systems, such as PAPR issues and channel quality adaptation. To mitigate the high PAPR problem, \cite{shao2022semantic} proposed PAPR reduction methods. They demonstrated a trade-off between performance and PAPR in semantic communication models, as reducing PAPR limits the representation power of the symbols. However, their evaluation was limited to AWGN simulations and showed only degraded performance with PAPR-restricted models, without demonstrating actual improvements in the nonlinear PA region.

Some prior work has explored adaptive encoding and decoding techniques based on given signal-to-noise ratio (SNR) values~\cite{xu2021wireless, ding2024adaptive, park2024joint}. These studies focus on training a single neural network to handle various SNRs, in contrast to typical semantic communications models which train separate neural networks for different target SNR values. These approaches, however, still assume a uniform SNR level for all symbols and do not account for varying SNRs for individual symbols.
Furthermore, in complex-valued I/Q symbol-based semantic communication systems, these approaches have shown slightly inferior performance compared to using an ensemble of neural networks trained separately for each SNR level~\cite{xu2021wireless}. In this article, to improve performance and align with prior testbed setups~\cite{yoo2022real, yoo2023role, liu2022real}, we adopt an ensemble of neural networks trained for each SNR level.

The authors in \cite{wu2024deep} proposed a neural-network-augmented MIMO semantic system combining zero forcing (ZF) equalization and neural network-based symbol refinement. However, their evaluation is confined to theoretical Rayleigh-fading channels. In contrast, we deploy a full MIMO-OFDM prototype under real-world impairments, such as PA nonlinearity and frequency-selective SNR variations. Rather than proposing novel MIMO processing techniques, our objective is to analyze practical challenges via prototype, including the simulation-prototype performance discrepancies reported in \cite{yoo2023role,liu2022real,ding2024adaptive}, and to derive actionable insights for future system designs.




\section{Challenges and Applied Solutions}
\label{sec:challenges}
Transmitting I/Q symbols from neural networks via wireless channels involves various challenges, primarily due to hardware constraints related to analog-to-digital converters (ADCs), digital-to-analog converters (DACs), and PAs. This section explores the challenges associated with neural network-based symbol generation and specifically how we address them in our prototype implementations.

\subsection{Fixed Point Representations}
Neural networks typically use 32-bit floating-point numbers in graphics processing units (GPUs). In wireless transmission, however, converting these digital symbols into voltages suitable for an antenna requires a fixed-point format. This is because the ADC and DAC of the transceiver internally use a fixed-point format, which limits the signal's dynamic range and resolution.

For instance, the ADC and DAC in the NI USRP-2953R SDR device support peak-to-peak voltage values ($V_{\mathrm{pp}}$) of 1, restricting signal amplitudes to the range $[-1, 1]$. If the neural network generates symbols exceeding this range, signal clipping may occur. To address this issue in our prototype implementations, we apply a normalization factor $N$ to scale the generated symbols, ensuring they remain within the allowable amplitude limits. 
The normalization process involves scaling the symbol by the constant $N$ and then applying a clipping function to restrict the values to the range $[-1, 1]$. The resulting normalized and clipped symbol is then ready for transmission.

Fixed-point representations in transceiver hardware also typically offer lower resolutions (commonly 14 or 16 bits) than floating-point formats, potentially leading to quantization errors. However, we conclude that the effects of symbol quantization due to fixed-point representations may be negligible for common ADC/DAC resolutions, as we observe no performance difference across a range of quantization levels. This is because the quantization noise level is significantly lower than the typical noise level in wireless channels. For instance, with a 14-bit fixed-point representation, the signal-to-quantization-noise ratio (SQNR) is approximately 22 dB when the signal uses only a quarter of the available range (e.g., $[-0.25, 0.25]$). Since typical SNR in communication systems ranges from 0 to 20 dB, we do not apply additional methods to mitigate quantization effects.

\textit{\textbf{Remark 1:} Semantic symbols should be properly normalized to prevent signal clipping. Quantization noise should be considered but may be negligible if the SQNR is sufficiently high.}

\subsection{Varying SNR Across Frequencies and Streams}
In practical OFDM systems, channel gain and noise vary across frequencies due to frequency-selective fading and hardware imperfections. In MIMO systems, different spatial streams can experience varying SNR levels due to antenna patterns, MIMO equalization, and other factors. As a result, symbols transmitted over different subcarriers and streams can experience different SNRs, leading to non-uniform and possibly multimodal symbol error distributions.

\begin{figure}[t]
    \centering{
        \includegraphics[width=0.95\columnwidth]{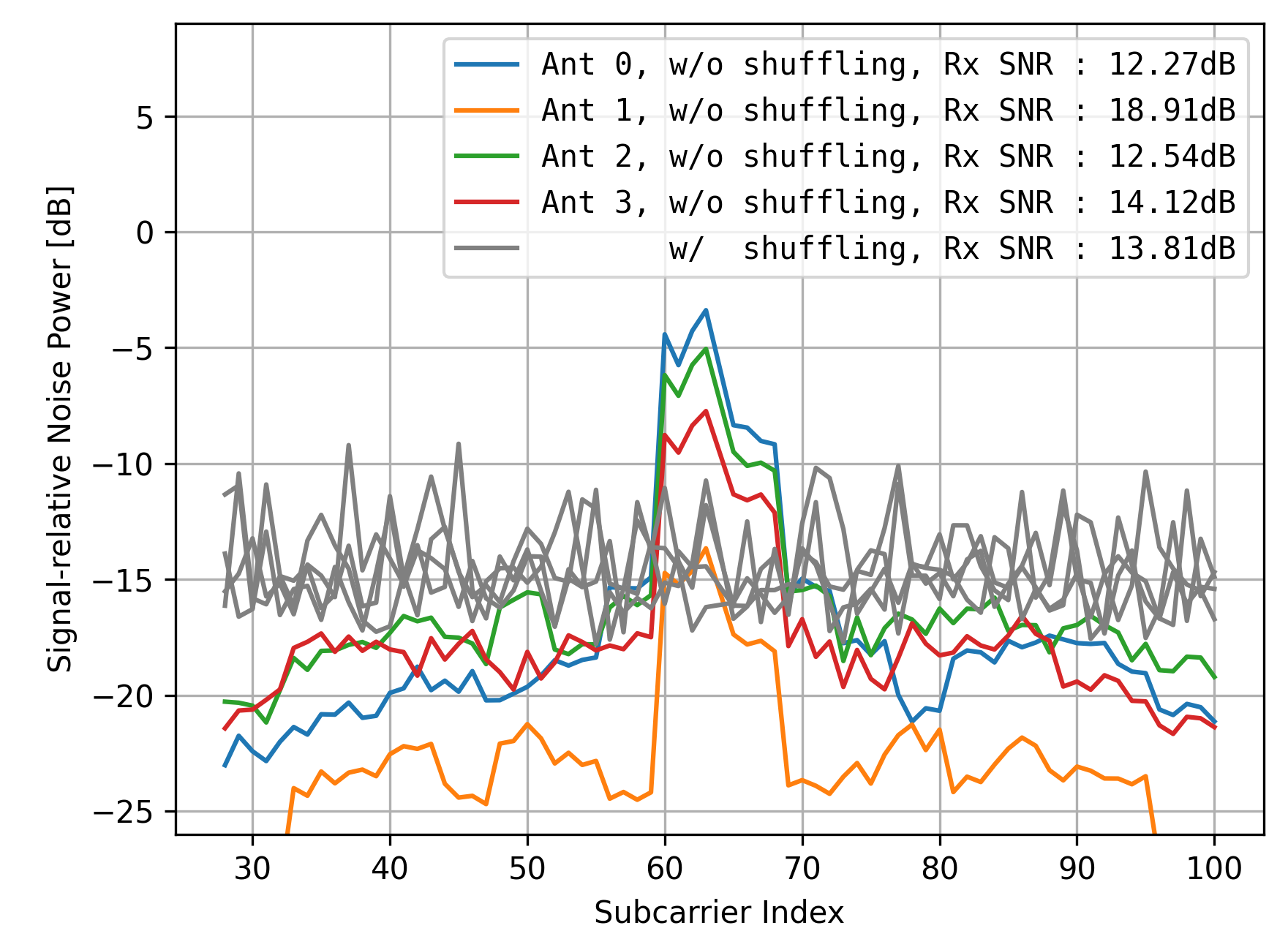}
    }
    \caption{Error plot from the wireless prototype, illustrating varying noise levels across subcarriers and data streams. Shuffling allocates symbols randomly across subcarriers, time slots, and streams to prevent concentration of poor channels into specific ranges of symbols. In the shuffled plot, the subcarrier index indicates where symbols would be allocated without shuffling.}
    \label{fig:err_spectrum}
\end{figure}

Fig.~\ref{fig:err_spectrum} shows the error of a signal across the subcarrier index and MIMO data streams transmitted in our wireless prototype (colored) and the error with shuffled symbols before transmission (gray). These error variations result in varying SNR conditions for symbols that are simultaneously fed into the semantic encoder. We hypothesize that these SNR variations cause the inconsistencies between simulation results and real-world data reported in prior work.

To support this hypothesis, inspired by the results in Fig.~\ref{fig:err_spectrum}, we mitigate this variance by shuffling the symbol sequence before mapping the symbols to resource blocks, similar to interleaving in conventional communication systems. This ensures that every adjacent symbol experiences the same average level of channel and noise, bringing the practical system closer to the AWGN assumption. By doing so, we conclude that the performance gap between simulations and prototype results reported in previous works~\cite{yoo2023role, xu2021wireless} is mainly due to these phenomena. We also show that near-AWGN performance can be achieved if those channel variations are conscientiously addressed.

\textit{\textbf{Remark 2:} The discrepancies between simulations and prototype results reported in prior literature are primarily due to variations in SNR conditions across symbols. Near-theoretical performance can be achieved if these variations are carefully accounted.}

\subsection{PAPR and PA Nonlinearity}

\begin{figure}[t]
    \centering
    \includegraphics[width=\columnwidth]{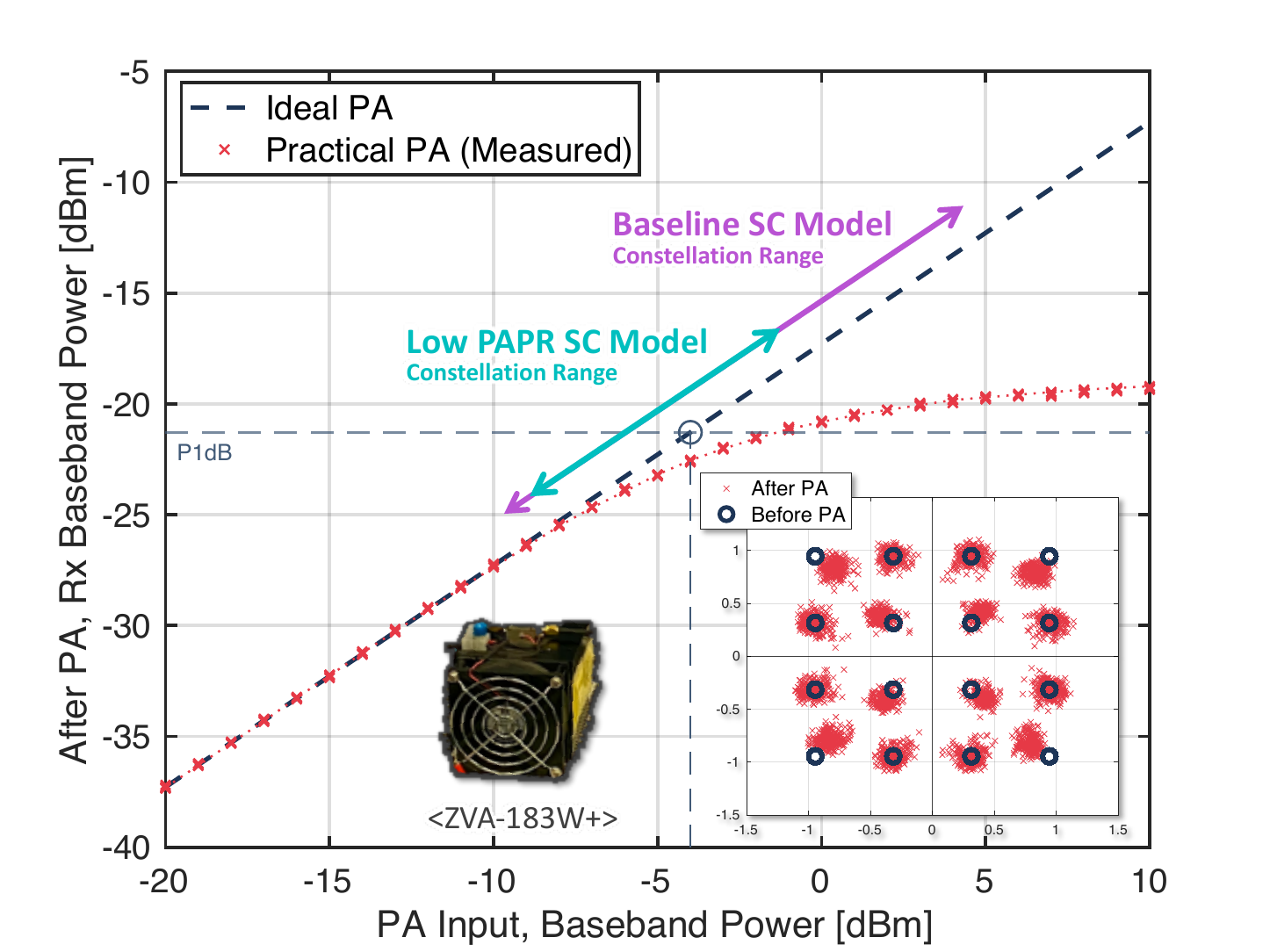}
    \caption{Nonlinear input-output power relationship of the power amplifier. Arrows mark the symbol peak power of each model at -5 dBm PA input.}
    \label{fig:pa_nonlinearity}
  \end{figure}

In actual wireless transmission systems, significant variations in symbol power levels can cause high PAPR and nonlinear signal distortion. This distortion occurs because the gain of a PA varies with the amplitude of the input signal, as shown in Fig.~\ref{fig:pa_nonlinearity}. This issue is exacerbated in semantic communications systems, where typical 95th percentile PAPR values are around 9.8~dB, compared to 8.5~dB for 16-QAM in 72-subcarrier OFDM environments.

To mitigate this issue, as in prior work~\cite{shao2022semantic, liu2023comprehensive}, we used a loss-based method to control the PAPR of the symbols. We introduced a PAPR penalty term in addition to the original mean squared error (MSE) loss, with a multiplier $\lambda_{\text{PAPR}}$ for the PAPR term. Note that our purpose is not to propose a new PAPR algorithm, but to validate existing loss-based PAPR reduction techniques in the PA’s nonlinear region, which has not been empirically demonstrated in prior work.

\begin{figure}[t]
    \centering
    \includegraphics[width=0.95\columnwidth]{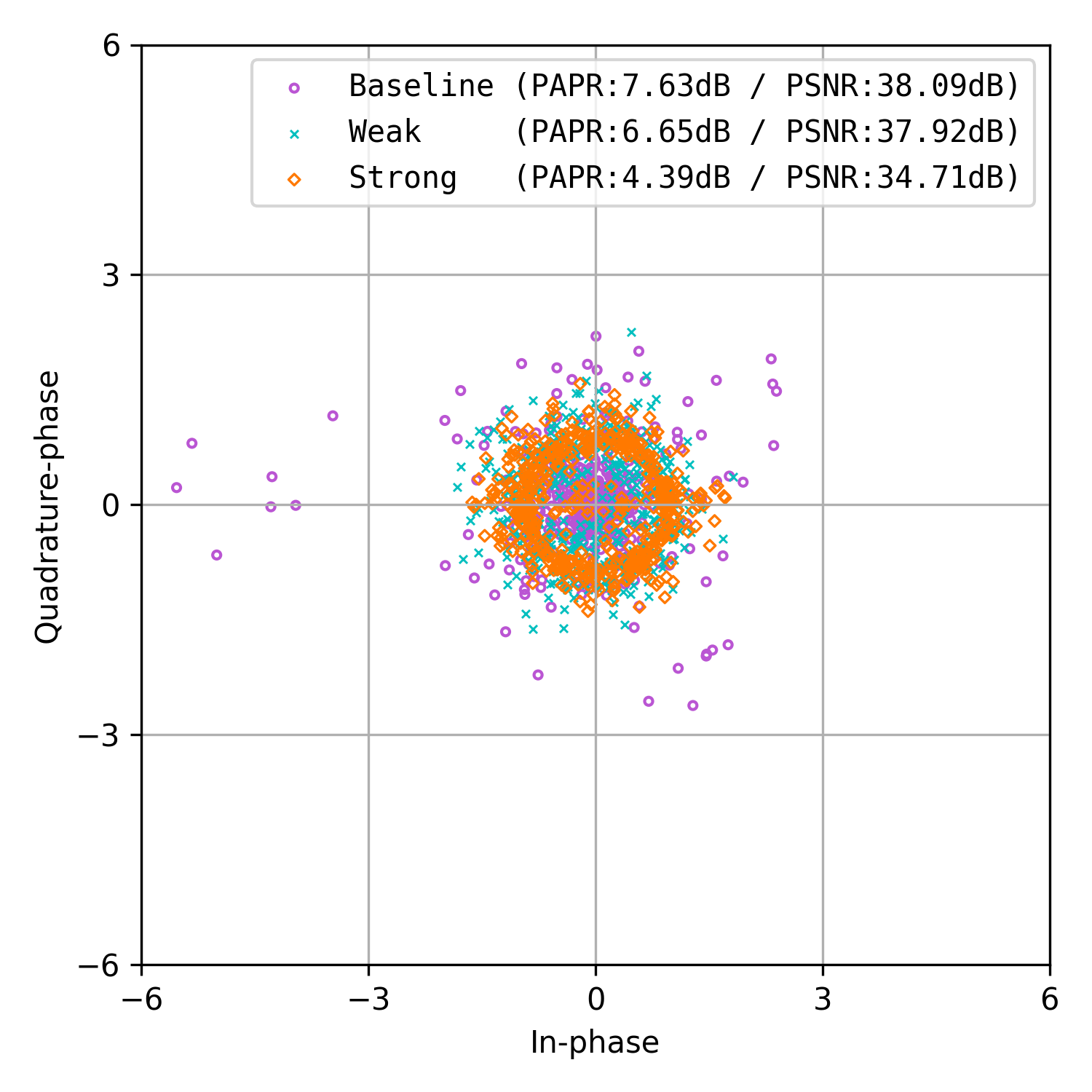}
    \caption{Constellation diagram from various PAPR-restricted semantic models with corresponding PSNR and PAPR values of the OFDM waveform with 72 subcarriers. Baseline, weak, and strong indicate the level of PAPR regulation of the semantic model, where the model is trained with PAPR loss weights of 0, 1/32768, and 1/4096, respectively.}
    \label{fig:constellations_result}
\end{figure}

Fig.~\ref{fig:constellations_result} illustrates the effect of PAPR reduction on the symbol constellation and the PA's nonlinear behavior. In Fig.~\ref{fig:constellations_result}, three constellation diagrams are shown for different PAPR constraints: baseline (no PAPR constraint), weak PAPR regulation, and strong PAPR regulation. We used $\lambda$ values for the PAPR loss of 0, $\frac{1}{32768}$, and $\frac{1}{4096}$, respectively. The baseline model displays widely dispersed symbols with a high PAPR of approximately 7.6~dB, increasing susceptibility to distortion. Introducing weak PAPR regulation tightens the symbol clustering, reducing PAPR to around 6.6~dB and slightly lowering the peak signal-to-noise ratio (PSNR). Here, PSNR quantifies the reconstruction error between the original and compressed/received image on a dB scale, with higher values indicating better image quality.
Strong PAPR regulation further confines the symbols near the origin, lowering PAPR to about 4.3 dB, albeit at the cost of worse PSNR and reduced symbol representation power. Due to this reduced representation power, in simulated channels without PA nonlinearity, this additional loss results in performance degradation due to the reduced representation power of the encoded symbols~\cite{shao2022semantic}.

In a real wireless setup with PA nonlinearity, however, the model with reduced PAPR can outperform the baseline model despite the performance degradation from the PAPR reduction. Higher PAPR symbols from the baseline model extend into the PA's nonlinear region, causing significant distortion. In contrast, PAPR-reduced models remain within the linear operating range of the PA, minimizing distortion and enhancing signal quality.

\textit{\textbf{Remark 3:} Semantic models have higher PAPR. PAPR-reduced models perform better in nonlinear power regions despite the performance tradeoff.}

\section{Prototype Implementations}
\label{sec:implementations}

We adopt a hybrid approach by training the semantic encoder-decoder under an AWGN model while using ZF channel estimation at run time. This setup ensures generalization across diverse deployments, allowing us to provide general insights for future semantic system designs.
We implemented a wireless semantic communications prototype for image reconstruction, integrating neural network processing on an Nvidia RTX 2080 Ti GPU server and wireless transmission via an NI USRP-2953R SDR. Fig.~\ref{fig:testbed} illustrates the MIMO-OFDM setup.

Following~\cite{yoo2023role}, our encoder employs a CNN-Transformer architecture to convert each $32\times32\times3$ color image patch into 512 complex symbols, normalized with $N=3$. This architecture combines CNN's local feature extraction with the Transformer's long-range modeling. The system processes larger images by dividing them into patches, with each patch serving as a basic processing unit. The symbols are modulated using OFDM with 128-point FFT and 72 active subcarriers, aligned with LTE's 1.4~MHz bandwidth (15~kHz spacing, 66.67~$\mu$s symbol). DC subcarriers are excluded to avoid RF bias distortion. Zadoff-Chu sequences enable synchronization, while 4-QAM pilots (similar to LTE DMRS) are inserted every 7 OFDM symbols on alternating subcarriers per antenna. Channel estimation uses linear interpolation in time and nearest-neighbor interpolation in frequency.

We employ open-loop spatial multiplexing with ZF receivers in 2$\times$2 MIMO and 4$\times$4 MIMO configurations. We believe that our approach and insights are extendable to larger antenna arrays for higher throughput, albeit with increased complexity. All antennas were positioned in a line-of-sight (LoS) environment with limited scattering, as shown in Fig.~\ref{fig:testbed}. We use actual over-the-air wireless channels measured in a static indoor environment, and Tx/Rx separation was about 1 m. Although we omit precoding due to its complexity in this context, our open-source code includes partial support for it. 
Baseband signals are transmitted at 2~GHz and received over a 1~Gbps Ethernet link. Tx/Rx gains vary from 0-31.5~dB. A Mini-Circuits ZVA-183W-S+ amplifier (28.6~dB gain at 15~V) introduces PA nonlinearity for relevant tests; otherwise, it is bypassed. Received signals are passed to the GPU server, where ZF combining and neural decoding reconstruct the final image.

\section{Experiments and Results}
\label{sec:results}

\subsection{Experiment Details}
We evaluate reconstruction quality using PSNR on $256\times256$ images formed by tiling 64 CIFAR-10 patches.
As a baseline, we use BPG (HEVC-based) compression, LDPC coding, and QAM modulation. For PSNR calculation, we use 4:4:4 chroma subsampling. Unlike our patch-based semantic system, BPG encodes the entire image at once, which may give it a slight PSNR advantage.

Semantic models are trained on AWGN channels at SNRs of $\{0, 5, 10, 15, 20\}$~dB using CIFAR-10 and Adam optimizer ($\text{lr} = 10^{-4}$). At test time, we select the model that matches the measured Rx SNR. The bandwidth ratio, defined as the ratio of complex symbols to pixels, was $1/6$, resulting in 512 symbols per $32\times32$ image patch. Processing a full $256 \times 256$ image required encoding 64 patches, totaling $64 \times 512 = 32,768$ symbols.

Our system outputs I/Q symbols directly without relying on conventional modulation. This enables end-to-end optimization, unlike~\cite{ding2024adaptive}, which employs traditional modulation and quantization.
For the BPG+LDPC+QAM configuration, we explored combinations of LDPC code rates from the set $\{(3072, 6144),\ (3072, 4608),\ (1536, 4608)\}$, corresponding to code rates of $1/2$, $2/3$, and $1/3$, respectively, along with QAM modulation orders of $\{4,\ 16,\ 64,\ 256\}$. We report the best performance across all combinations. The BPG quantization parameter is adjusted to fit the symbol budget defined by the code rate and modulation. Due to LDPC's block structure, some settings yield slightly longer symbol lengths than our semantic baseline.


\begin{figure*}[t]
    \centering
    \subfloat[]{%
        \includegraphics[width=0.49\textwidth]{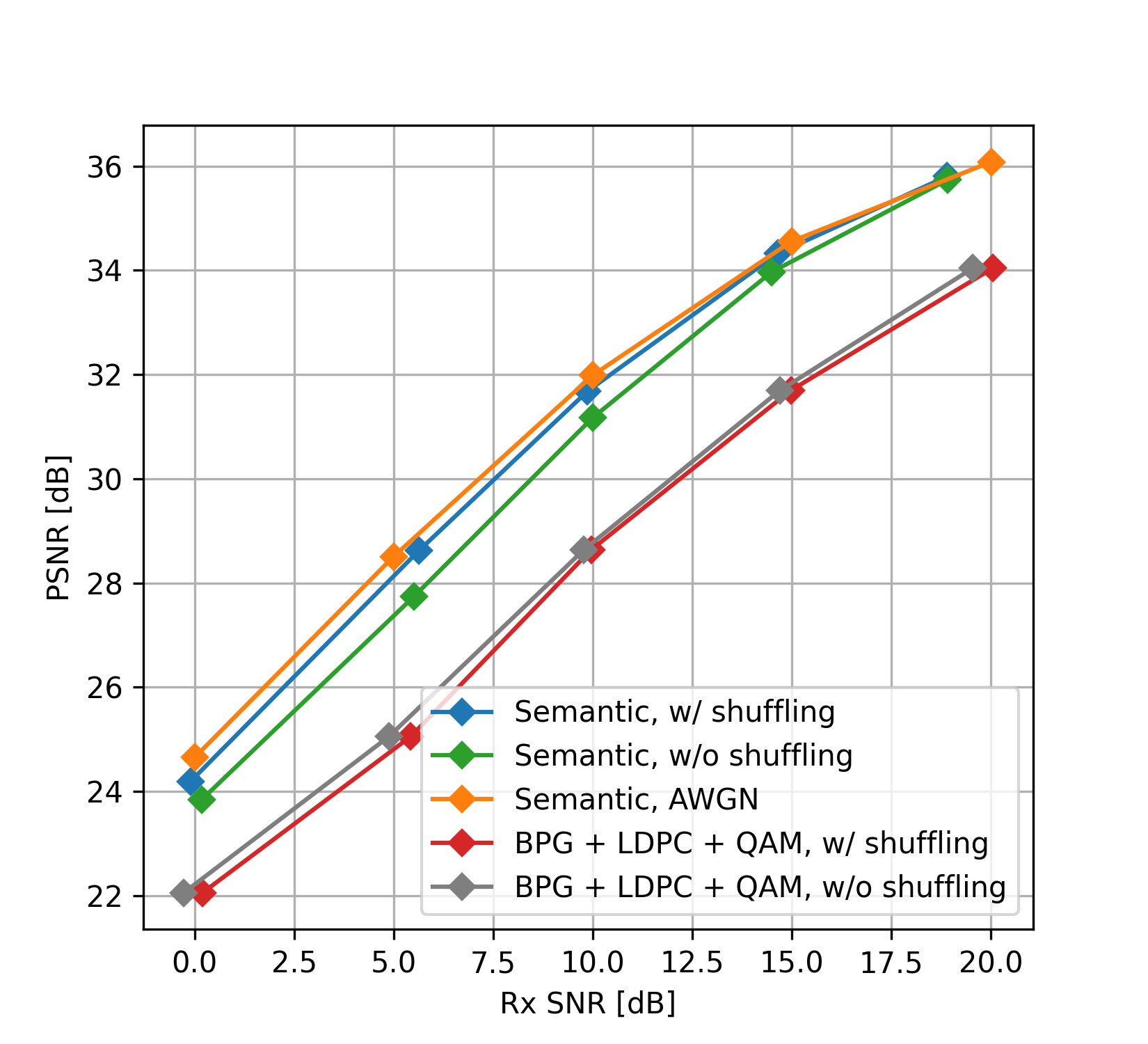}
        \label{fig:psnr_result}
    }
    \hfill
    \subfloat[]{%
    \includegraphics[width=0.49\textwidth]{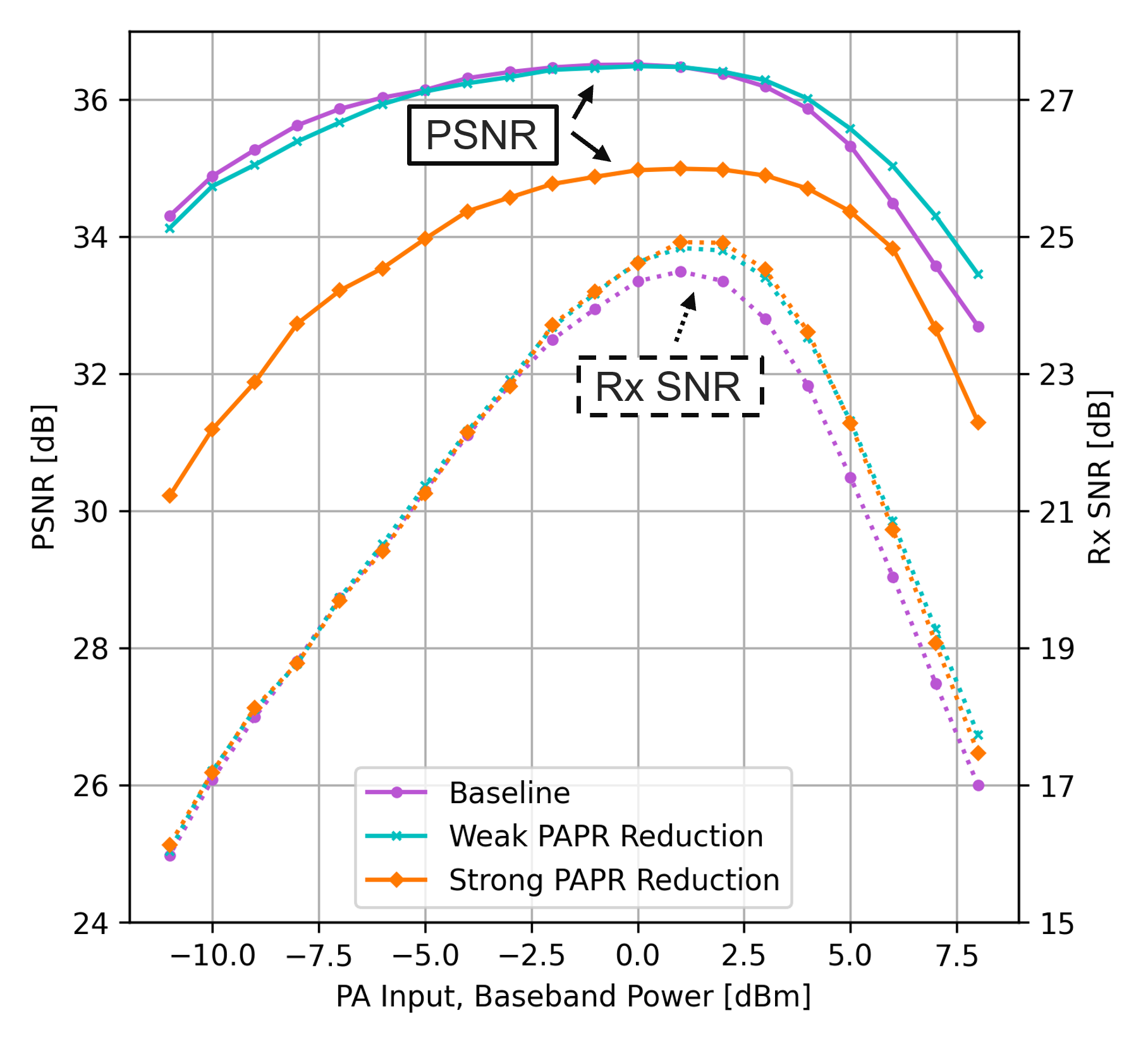}
        \label{fig:papr_tradeoff}
    }
    \caption{(a) Linear region result (MIMO-OFDM, Sim vs. OFDM-Semantic vs. OFDM-LDPC) for a $2\times2$ MIMO configuration. We also observed similar trends for SSIM (not shown in the article due to page limit). (b) Nonlinear region result (Sim vs. low-PAPR vs. baseline models). In (b), the solid line and dotted line indicate the power-PSNR curve and the power-Rx SNR curve for a $2\times2$ configuration, respectively.}
\end{figure*}

\subsection{Transmission Performance in the Linear Power Region}
To assess the system's performance, we plot the decoded image quality (PSNR) against the receiver SNR (Rx SNR), defined as the power ratio between the transmitted symbol and the errors arising from sources such as channel noise, imperfect channel estimation, and PA nonlinearity. Here, the Rx SNR is computed based on the error vector magnitude (EVM). Note that in Fig.~\ref{fig:psnr_result}, the reported SNR values may slightly deviate from the target values (e.g., 5~dB) due to measurement limitations. We carefully adjusted the transmit power to approximate the target SNR and report the average Rx SNR over 15 independent transmissions per point.


Since the errors are not perfectly Gaussian distributed, the actual performance of the prototype would be equal to or less than the simulated AWGN channel at the same Rx SNR. Thus, the gap between the simulated and prototype results at the same Rx SNR reflects the non-Gaussian nature of the symbol error.

Fig.~\ref{fig:psnr_result} compares the reconstructed image quality between semantic and conventional encoder-decoder systems, evaluated on both prototype measurements and simulations. For each target Rx SNR (set by adjusting the transmit power), we conducted 15 independent experiments and report the resulting mean Rx SNR and PSNR values. The semantic communications systems outperform conventional BPG+LDPC+QAM systems across all regions in both simulated and prototype results.  Moreover, as we hypothesized in the previous sections, the prototype results with symbol shuffling closely align with those from the simulated AWGN channel, while a gap that expands with decreasing Rx SNR is observed without shuffling. This finding aligns with previous studies~\cite{yoo2023role,liu2022real}, indicating that the gap identified in earlier research is primarily due to noise level variations between subcarriers and streams. 

Notably, we observe no performance difference between BPG+LDPC+QAM results with and without shuffling. This is likely because LDPC already performs interleaving within its block size, typically spanning one or two images, yielding a similar effect. Moreover, conventional BPG+LDPC+QAM systems benefit from quantized modulations and error correction, which offer robustness to noise variation. In contrast, semantic communication systems directly decode the received symbols, making them more sensitive to variations in the error distribution.

These prototype results provide insight into how channel variations and resource allocations affect the performance of semantic communications systems in real environments, a topic not extensively covered in existing literature. They also validate the need for channel-adaptive semantic communications models, which provide actual channel information along with the symbols to the neural network for adaptive decoding. This approach is analogous to traditional communication systems, where modulation and coding scheme (MCS) information for each subcarrier or stream is necessary for optimal performance. In this article, for simplicity, we do not adopt adaptive encoding and decoding strategies, similar to prior works such as~\cite{yoo2022real,yoo2023role,liu2022real}. Nevertheless, we believe that the insights gained from our current setup remain applicable and can be extended to those adaptive models.

\subsection{Transmission Performance Based on PAPR Values in the Nonlinear Power Region}

To assess the reduction in nonlinearity due to decreased PAPR, we transmitted symbols from both the baseline and the low PAPR model, trained at 20~dB SNR, through a highly nonlinear PA region. We did not apply random shuffling of symbols in these experiments, as it affected symbol PAPR values.

As shown in Fig.~\ref{fig:papr_tradeoff}, both PSNR and Rx SNR increase with input power up to 0~dBm, then decline due to the distortion caused by PA nonlinearity. In this high-power, nonlinear region, models that incorporate PAPR reduction techniques exhibit approximately a 1~dB improvement in Rx SNR. This enhancement is attributed to the reduced PAPR, which minimizes the distortions introduced by the PA's nonlinear operation, thereby maintaining higher signal quality despite the increased input power.

Interestingly, while the weak PAPR reduction model outperforms the baseline in terms of PSNR in the 2.5 dBm or higher power region, the strong PAPR reduction model does not. This is due to the performance-PAPR tradeoff (see Fig.~\ref{fig:constellations_result}); excessive PAPR restriction leads to a performance drop that outweighs the benefits of reduced PAPR, resulting in overall lower PSNR despite similar Rx SNR levels. This indicates that PAPR reduction should be balanced to avoid diminishing returns.


\section{Limitations and Open Issues}
\subsection{Limitations}
\label{sec:limitations}
{
While our prototype provides insights into the practical challenges of semantic communication systems, several limitations remain:
}

{
\textbf{Environment Generalization:} Our experiments were conducted in line-of-sight environments with limited scattering. Performance may differ in non-line-of-sight or highly dynamic scenarios, which require further investigation.}

{
\textbf{Residual Performance Gap:} Although symbol shuffling mitigates many practical issues, a small gap between simulated and prototype results persists. Further analysis is needed to fully understand this discrepancy.
}

{
\textbf{AWGN-based Training and ZF Equalization:} To ensure generalizability and maintain consistency with prior testbed studies~\cite{yoo2022real,yoo2023role,liu2022real,ding2024adaptive}, we used models trained on AWGN channels and employed zero-forcing channel estimation at runtime.
However, this may not be optimal for all deployment scenarios.
Training models on real or site-specific channels, along with optimized resource allocation and precoding strategies, may improve performance but were beyond the scope of this study.
}

\subsection{Open Issues}
\label{sec:open_issues}

{Following the limitations discussed above, we highlight several open issues that must be addressed to advance the practical deployment of semantic communication systems.}

\textbf{Signal Processing Techniques for MIMO-OFDM Semantic Communications:}
We demonstrated that symbol mapping to resource blocks or streams affects system performance. While we employed basic symbol shuffling and ZF-based channel equalization, advanced techniques like MIMO precoding or optimized resource allocation could further enhance performance. Although MIMO precoding offers limited benefits in our setup due to few antennas and feedback complexity, it may be advantageous in larger MIMO configurations.

\textbf{Channel-Adaptive Semantic Communications:}
{
While several recent works have explored channel-adaptive semantic communications~\cite{xu2021wireless, ding2024adaptive, park2024joint}, fine-grained adaptation to varying channel conditions across subcarriers and streams remains underexplored. Future research should investigate strategies such as allocating semantic symbols with different importance levels to channel resources with varying quality, in order to further optimize system performance.}

\textbf{Real Channel-based Training of Semantic Communications Models:}
Training semantic communication models with actual channel measurements can enhance reliability by capturing site-specific characteristics such as obstructions and interference patterns. However, this may lead to overfitting and reduce generalizability. Therefore, balancing adaptation to real channels with robustness across diverse scenarios is essential.

\textbf{Adaptation to Modern Communication Systems:}
For integration into future wireless technologies, semantic communications must be validated with advanced techniques like massive MIMO and full-duplex (FD) systems. Additionally, testing in challenging scenarios such as high mobility or urban environments and dynamically adapting to varying channels is crucial~\cite{xu2021wireless}.

\textbf{Efficient Semantic Communications Systems:}
Semantic communications often utilize complex neural networks that introduce high latency, hindering real-time functionality. Developing lightweight architectures and applying model compression and quantization techniques can significantly reduce latency and computational demands, making systems more practical for real-time applications.

\textbf{Semantic Systems for Videos or Ultra-High-Resolution Images:}
Extending semantic communication to high-resolution images and video~\cite{tung2021deepwive} remains challenging due to the large data volume and stringent real-time requirements. Developing efficient encoding and temporal modeling techniques that account for wireless channel dynamics is an important open problem.


\section{Conclusion}
\label{sec:conclusion}

We developed a MIMO-OFDM-based wireless semantic communications system to investigate the impact of channel variations and resource allocations on real-world performance. Our findings indicate that the performance gap between simulations and prototypes is primarily due to varying SNR levels across symbols. By implementing techniques such as symbol shuffling, our prototype achieved performance comparable to simulations, highlighting the need for channel-adaptive semantic communication models. Additionally, we demonstrated that PAPR reduction techniques can enhance performance by mitigating PA nonlinearity effects, despite the tradeoff between PSNR and PAPR. Our source code and hardware setup are publicly available for further validation. While this study focused on reconstruction-based semantic communication, the methods and insights can be extended to other tasks like classification and detection. We believe these findings will drive future innovations and promote the broader adoption of semantic communication systems in practical applications.


\bibliographystyle{IEEEtran}
\bibliography{ref}

\begin{IEEEbiographynophoto}\noindent\textbf{Hanju Yoo} (Student Member, IEEE) received the B.S. degree (summa cum laude) from the School of Integrated Technology, Yonsei University, South Korea, in 2021, where he is currently pursuing his Ph.D. degree. His research interests include semantic communications, information theory for deep learning, and wireless system prototyping.
\end{IEEEbiographynophoto}

\begin{IEEEbiographynophoto}\noindent\textbf{Dongha Choi} (Student Member, IEEE) received the B.S. degree from the School of Integrated Technology, Yonsei University, South Korea, in 2023, where he is currently pursuing his Ph.D. degree. His research interests include AI-based communications and information theory for deep learning.
\end{IEEEbiographynophoto}

\begin{IEEEbiographynophoto}\noindent\textbf{Yonghwi Kim} (Member, IEEE) received his B.S. and Ph.D. degrees in School Integrated Technology from Yonsei University, Korea. His research focuses on machine learning-based signal processing and integrated sensing and communications (ISAC) for B5G/6G wireless systems, with an emphasis on full-duplex (FD) technologies.
\end{IEEEbiographynophoto}

\begin{IEEEbiographynophoto}\noindent\textbf{Yoontae Kim} (Student Member, IEEE) received the B.S. degree from the School of Integrated Technology, Yonsei University, in 2022, where he is currently pursuing the Ph.D. degree. His research interests include prototyping for next-generation wireless communication networks, with a focus on MIMO, full-duplex radios, semantic communications, and fluid antenna systems.
\end{IEEEbiographynophoto}

\begin{IEEEbiographynophoto}\noindent\textbf{Songkuk Kim} (Member, IEEE) received the Ph.D. degree in computer science from the University of Michigan, Ann Arbor, MI, USA, in 2005. From 2005 to 2007, he was a Research Staff with Xerox Research Center, and was with Google as a Software Engineer, until 2011. He is currently an Assistant Professor with the School of Integrated Technology, Yonsei University, South Korea. His research interests include machine learning, big data mining, and cloud computing.
\end{IEEEbiographynophoto}

\begin{IEEEbiographynophoto}\noindent\textbf{Chan-Byoung Chae} (Fellow, IEEE) is an Underwood Distinguished Professor and Lee Youn Jae Fellow (Endowed Chair Professor) with the School of Integrated Technology at Yonsei University, Seoul, South Korea. He is also the Chief Scientific Officer (CSO) of SensorView, Ltd. He is the author of \emph{Signal Processing Engineering: An Intuitive Approach} (Springer, 2026) and has co-authored more than 200 journal papers. He has received several awards from CES, IEEE Communications Society (ComSoc), IEEE Signal Processing Society (SPS), and IEEE Vehicular Technology Society (VTS). He is an elected member of the National Academy of Engineering of Korea.
\end{IEEEbiographynophoto}

\begin{IEEEbiographynophoto}\noindent\textbf{Robert W. Heath, Jr.} (Fellow, IEEE) is  the Charles Lee Powell Chair of Wireless Communications with the Department of Electrical and Computer Engineering, University of California at San Diego, CA, USA. He is also the President and the CEO of MIMO Wireless Inc. He has authored \emph{Introduction to Wireless Digital Communication} (Prentice Hall, 2017) and \emph{Digital Wireless Communication: Physical Layer Exploration Lab Using the NI USRP} (National Technology and Science Press, 2012) and coauthored \emph{Millimeter Wave Wireless Communications} (Prentice Hall, 2014) and \emph{Foundations of MIMO Communication} (Cambridge University Press, 2018).  He received the 2025 IEEE/RSE James Clerk Maxwell Medal. He is an elected member of the National Academy of Engineering in the USA.
\end{IEEEbiographynophoto}

\end{document}